\documentclass[referee]{aa} 
\usepackage{graphicx}
\usepackage{txfonts}
\begin{document}

   \title{Determining the equation of state of dark energy from
	  angular size of compact radio sources and 
	  X-ray gas mass fraction of galaxy clusters 
	 }   


   \author{Zong-Hong Zhu
	  \inst{1} 
          \and
	  Masa-Katsu Fujimoto
          \inst{1}
	  \and
	  Xiang-Tao He
	   \inst{2}
          }

   \offprints{Zong-Hong Zhu}

   \institute{National Astronomical Observatory,
	      2-21-1, Osawa, Mitaka, Tokyo 181-8588, Japan\\
	      \email{zong-hong.zhu@nao.ac.jp ~~fujimoto.masa-katsu@nao.ac.jp}
	  \and
	     Department of Astronomy, Beijing Normal University, 
	     Beijing 100875, China\\
	     \email{xthe@bnu.edu.cn}	
             }

   \date{Received 00 00, 0000; accepted 00 00, 0000}

   \abstract{
   Using recent measurements of angular size of high-$z$ milliarcsecond
     compact radio sources compiled by Gurvits, Kellermann \& Frey (1999) and 
     X-ray gas mass fraction of galaxy clusters published by Allen et al. 
     (2002,2003), we explore their bounds on the equation of
     state, $\omega_x \equiv p_x/\rho_x$, of the dark energy, 
     whose existence has been congruously suggested by various cosmological
     observations. 
   We relaxe the usual constraint $\omega_x \ge -1$, and find that combining
     the two databases yields a nontrivial lower bound on $\omega_x$.
   Under the assumption of a flat universe, we obtain a bound 
     $-2.22 < \omega_x < -0.62$ at 95.4\% confidence level.
   The 95.4\% confidence bound goes to $-1 \le \omega_x < -0.60$ when the
     constraint $\omega_x \ge -1$ is imposed.
   }

    \authorrunning{Zhu, Z.-H., Fujimoto, M.-K. \& He, X.-T.}

    \titlerunning{Determining $\omega_x$ from 
		$\Theta$-$z$ and $f_{\rm gas}$ data}

   \maketitle
%

\section{Introduction}

One of the most remarkable cosmological findings of recent years is,
  in additional to the cold dark matter (CDM), the existence of a
  component of dark energy (DE) with negative pressure in our universe.
It is motivated to explain the acceleration of the universe discovered
  by distant type Ia supernova (SNeIa) observations
  (Perlmutter et al. 1998, 1999; Riess et al. 1998, 2001),
  and to offset the deficiency of a flat universe, favoured by
  the measurements of the anisotropy of CMB
        (de Bernardis et al. 2000;
        Balbi et al. 2000,
        Durrer et al. 2003;
        Bennett et al. 2003;
        Spergel et al. 2003),
  but with a subcritical matter density parameter $\Omega_m \sim 0.3$,
  obtained from dynamical estimates or X-ray and gravitational lensing 
  observations of clusters of galaxies(for a recent summary, see Turner 2002).
While a cosmological constant with $p_{\Lambda} = - \rho_{\Lambda}$ 
  is the simplest candidate for DE,
  it suffers from the difficulties
  in understanding of the observed value in the framework of modern  quantum
  field theory (Weinberg 1989; Carroll et al. 1992) and
  the ``coincidence problem'', the issue of
  explaining the initial conditions necessary to yield the near-coincidence
  of the densities of matter and the cosmological constant component today.
In this case, quintessence (a dynamical form of DE with
  generally negative pressure) has been invoked
	(Ratra and Peebles 1988;
	Wetterich 1988;
	Caldwell, Dave and Steinhardt 1998;
	Zlatev, Wang and Steinhardt 1998).
One of the important characteristics of quintessence models is that
  their equation of state, $\omega_x \equiv p_x/\rho_x$, vary with cosmic time
  whilst a cosmological constant remains a constant $\omega_{\Lambda}=-1$. 
Determination of values of $\omega_x$ and its possible cosmic evolution plays
  a central role to distinguish various DE models.
Such a challenging has triggered off a wave of interest aiming to constrain 
  $\omega_x$ using various cosmological databases,
  such as SNeIa 
	(Garnavich et al. 1998; 
	Tonry et al. 2003; 
	Barris et al. 2003;
	Knop et al. 2003;
	Zhu and Fujimoto 2003);
    old high redshift objects (Lima and Alcaniz 2000a);	
    angular size of compact radio sources (Lima and Alcaniz 2002);
    gravitational lensing
	(Chae et al. 2002;
	Sereno 2002;
	Dev, Jain and Mahajan 2003; 
	Huterer and Ma 2003);
    SNeIa plus Large Scale Structure (LSS) (Perlmutter, Turner \& White 1999);
    SNeIa plus gravitational lensing (Waga and miceli 1999);
    SNeIa plus X-ray galaxy clusters (Schuecker et al. 2003);
    CMB plus SNeIa 
	(Efstathiou 1999;
	Bean and Melchiorri 2002;
	Hannestad and M\"ortsell 2002;
	Melchiorri et al. 2003);
    CMB plus stellar ages (Jimenez et al. 2003);
    and combinations of various databases (Kujat et al. 2002).
Other potential methods for the determination of $\omega_x$ have also widely
  discussed in literatures, such as
    the proposed {\it SNAP} satellite\footnote{{\it SNAP} home page, 
   	http://snap.lbl.gov}
	(Huterer and Turner 1999;
	Weller and Albrecht 2001;
	Weller and Albrecht 2002);
    advanced gravitational wave detectors
	(Zhu, Fujimoto and Tatsumi 2001;
	Biesiada 2001);
    future SZ galaxy cluster surveys (Haiman, Mohr and Holder 2001);
    and gamma ray bursts
	(Choubey and King 2003;
	Takahashi et al. 2003).


In this work, we shall consider the observational constraints on the DE
  equation of state parameterized by a redshift independent 
  pressure-to-density ratio $\omega_x$ arising from
  the latest observations of angular size of high-$z$ milliarcsecond
  compact radio sources compiled by Gurvits, Kellermann \& Frey (1999)
  and the X-ray gas mass fraction data of clusters of galaxies 
	published by Allen et al. (2002, 2003).
The basics of a constant $\omega_x$ assumption are two folds:
  on the one hand, the angular diameter distance $D^A$ used in this work is
  not sensitive to variations of $\omega_x$ with redshift because it depends
  on $\omega_x$ through multiple integrals
	(Maor et al. 2001;
	Maor et al. 2002;
	Wasserman 2002);
  on the other hand, for a wide class of quintessence models (particularly, 
  those with tracking solutions), both of $\Omega_x$ and $\omega_x$ vary 
  very slowly (Zlatev et al. 1999; Steinhardt et al. 1999; Efstathiou 1999),
  and an effective equation of state, 
  $\omega_{\rm eff} \sim \int \omega_x(z) \Omega_x(z) dz / \int \Omega_x(z) dz$
  is a good approximation for analysis (Wang et al. 2000).
We relaxe the usual constraint $\omega_x \ge -1$, because recent years there
  have been several models which predict a DE component with 
  $\omega_x < -1$ 
	(Parker and Raval 1999;
	Schulz and White 2001;
	Caldwell 2002;
	Maor et al. 2002;
	Frampton 2003)
  and also we hope to explore its effects on the $\omega_x$ determination.
The confidence region on the ($\omega_x$, $\Omega_m$) plane obtained through
  a combined analysis of the two databases suggests
  $-2.22 < \omega_x < 0.62$ at 95.4\% confidence level,
  which goes to $-1 \le \omega_x < 0.60$ when the constraint $\omega_x \ge -1$
  is imposed.

The plan of the paper is as follows.
In the next section, we provide the bounds on $\omega_x$ from the angular 
  size-redshift data.
Constraints from the X-ray gas mass fraction of galaxy clusters are discussed
  in section~3.
Finally, we present a combined analysis, our concluding remarks 
  and discussion in section~4.
Throughout of the paper, we assume a flat universe which is suggested by 
  the measurements of the anisotropy of CMB and favoured by inflation scenario.


\section{Constraints from the angular size-redshift data} 

We begin by evaluating the angular diameter distance $D^A$ 
  as a function of redshift $z$.
The redshift dependent Hubble parameter can be written as
  $H(z) = H_0 E(z)$, where $H_0=100h\,$kms$^{-1}$Mpc$^{-1}$ is 
  the Hubble constant at the present time.
For a flat universe that contains (baryonic and cold dark) matter and 
  dark energy with a constant $\omega_x$ 
  (we ignore the radiation components in the universe that are not
   important for the cosmological tests considered in this work), 
  we get (Turner and White 1997; Chiba et al. 1997; Zhu 1998)
\begin{equation}
\label{eq:DA}
D^A(z; \Omega_m, \omega_x) =
                    {c \over H_0 }{1 \over {1+z}} \int_{0}^{z} {dz^{\prime}
                             \over E(z^{\prime};\Omega_m, \omega_x)}\; ,\;\;\;
E^2(z; \Omega_m, \omega_x) = \Omega_m (1+z)^3 + 
		(1-\Omega_m) (1+z)^{3(1+\omega_x)} .
\end{equation}

   \begin{figure*}
   \centering
   \includegraphics[width=8.0cm]{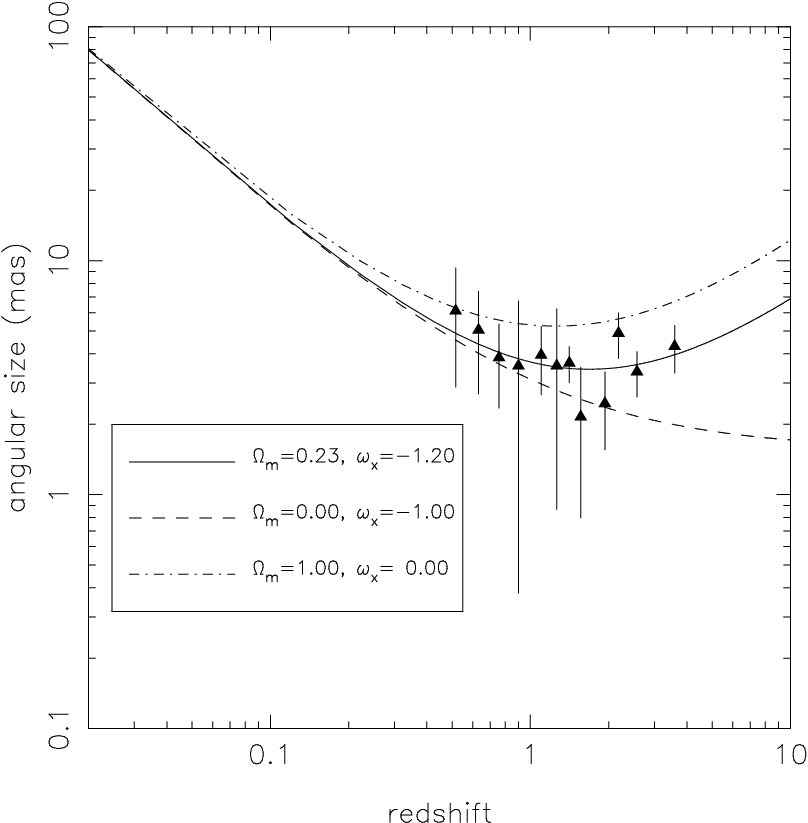}
   \caption{Diagram of angular size vs redshift data for 145 compact radio
        sources (binned into 12 bins) of Gurvits, Kellermann and Frey (1999).
	We assume the charateristic linear size $l = 22.64h^{-1}$pc for
	theoretical curves.
	The solid curve corresponds to our best fit with $\omega_x = -1.19$
	and $\Omega_m = 0.23$, while the dashed and dot-dashed curves 
	correspond to a $\Lambda$-dominated universe and the standard cold dark
	matter (SCDM) model respectively. 
        }
   \label{fig1:data}
   \end{figure*}

We first analyze the angular size-redshit data for milliarcsecond radio sources
  recently compiled by Gurvits, Kellermann and Frey (1999) to constrain
  $\omega_x$.
{
The basics of the angular size-redshit test in the context of dark energy was
  first discussed in a theoretical viewpoint by Lima and Alcaniz (2000b) 
  without using any database.
They also provide an analytical closed form which determines how the redshift
  $z_m$, at which the angular size takes its minimal value, depends on
  $\omega_x$.
Later on, using the same database compiled by Gurvits, Kellermann and Frey 
  (1999), Lima and Alcaniz (2002) obtained $\Omega_m \sim 0.2$ and 
  $\omega_x \sim -1.$.
}
A distinguishing characteristic of our analysis is that the usual constraints
  $\omega_x \ge -1$ is relaxed.
This database shown in Figure~1 is 145 sources distributed into twelve 
  redshift bins with about the same number of sources per bin.
The lowest and highest redshift bins are centered at redshifts $z=0.52$ and
  $z=3.6$ respectively.
We determine the model parameters $\omega_x$ and $\Omega_m$ through a $\chi^{2}$
  minimization method.
The range of $\omega_x$ spans the interval [-3,0] in steps of 0.01, while the
  range of $\Omega_m$ spans the interval [0, 1] also in steps of 0.01.
\begin{equation}
\label{eq:chi2}
\chi^{2}(l; \Omega_m, \omega_x) =
  \sum_{i}^{}{\frac{\left[\theta(z_{i}; l; \Omega_m, \omega_x)
     - \theta_{oi}\right]^{2}}{\sigma_{i}^{2}}},
\end{equation}
where $\theta(z_{i}; \Omega_m, \omega_x) = l/D^A$ is the angle subtended by an object
  of proper length $l$ transverse to the line of sight and $\theta_{oi}$ is
  the observed values of the angular size with errors $\sigma_{i}$ of the $i$th
  bin in the sample.
The summation is over all 12 observational data points.

As pointed out by the authors of previous analyses on this database
	(Gurvits, Kellermann and Frey 1999;
	Vishwakarma 2001;
	Alcaniz 2002;
	Zhu and Fujimoto 2002;
	Jain, Dev and Alcaniz 2003;
	Chen and Ratra 2003),
  when one use the angular size data to constrain the
  cosmological parameters, the results will be strongly dependent on the
  characteristic length $l$.
Therefore, instead of assuming a specific value for $l$, we have worked on
  the interval $l = 15h^{-1} - 30h^{-1}$pc.
In order to make the analysis independent of the choice of the characteristic
  length $l$, we also minimize equation~(2) for $l$, $\omega_x$ and
  $\Omega_m$ simultaneously, which gives
  $l=22.64h^{-1}$pc, $\omega_x = -1.19$ and $\Omega_m = 0.23$ 
  as the best fit.
Figure~2 displays the 68.3\% and 95.4\% confidence level contours in the 
  ($\Omega_m$, $\omega_x$) plane using the lower shaded and the lower plus 
  darker shaded areas respectively.
It is clear from the figure hat $\omega_x$ is poorly constrained from
  the angular size-redshift data alone, which only gives $\omega_x < -0.32$
  at 95.4\% confidence level.
However, as we shall see in Sec.4, when we combine this test with the X-ray
  gas mass fraction test, we could get fairly stringent constraints on both
  $\omega_x$ and $\Omega_m$.
   \begin{figure*}
   \centering
   \includegraphics[width=8.0cm]{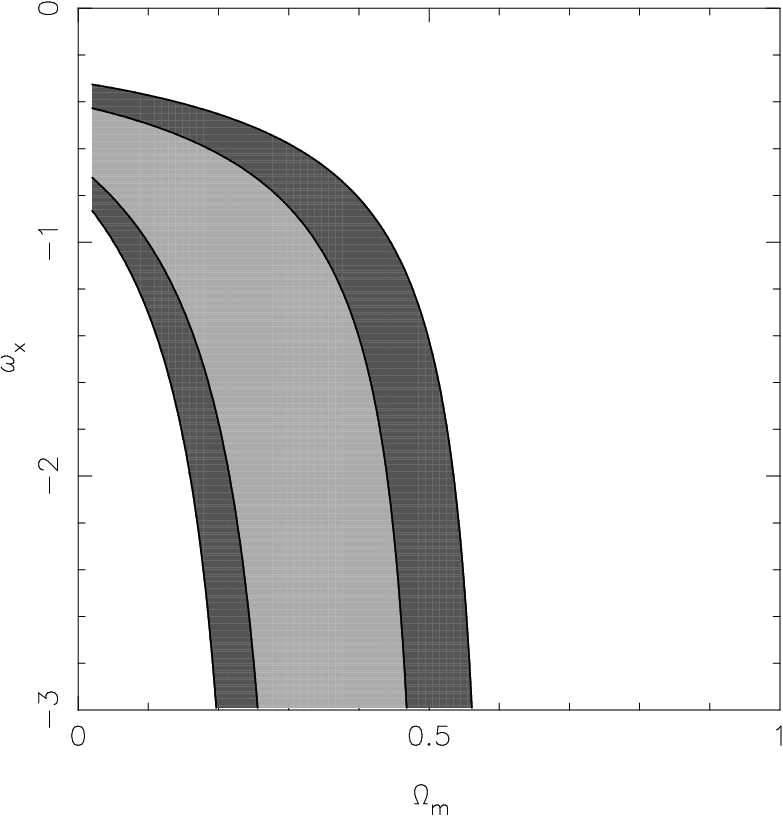}
   \caption{Confidence region plot of the best fit to the database of the
	angular size-redshift data compiled by Gurvits, Keller and Frey (1999)
	-- see the text for a detailed description of the method.
	The 68\% and 95\% confidence levels in the ($\Omega_m$, $\omega_x$) 
	plane are shown in lower shaded and lower $+$ darker shaded areas
	respectively.
           }
   \label{Fig_cont1}
    \end{figure*}
%


\section{Constraints from the galaxy clusters X-ray data}

Clusters of galaxies are the largest virialized systems in the universe,
  and their masses can be estimated by X-ray and  optical observations,
  as well as gravitational lensing measurements.
A comparison of the gas mass fraction, 
  $f_{\rm gas} = M_{\rm gas} / M_{\rm tot}$,
  as inferred from X-ray observations of clusters of galaxies to the cosmic
  baryon fraction can provide a direct constraint on the density parameter
  of the universe $\Omega_m$ (White et. al. 1993).
Moreover, assuming the gas mass fraction is constant in cosmic time,
  Sasaki (1996) show that the $f_{\rm gas}$ data of clusters of galaxies
  at different redshifts also provide an efficient way to constrain other
  cosmological parameters decribing the geometry of the universe.
This is based on the fact that the measured $f_{\rm gas}$ values for each
  cluster of galaxies depend on
  the assumed angular diameter distances to the sources as 
  $f_{\rm gas} \propto [D^A]^{3/2}$.
The ture, underlying cosmology should be the one which make these measured
  $f_{\rm gas}$ values to be invariant with redshift 
  (Sasaki 1996; Allen at al. 2003).

   \begin{figure*}
   \centering
   \includegraphics[width=8.0cm]{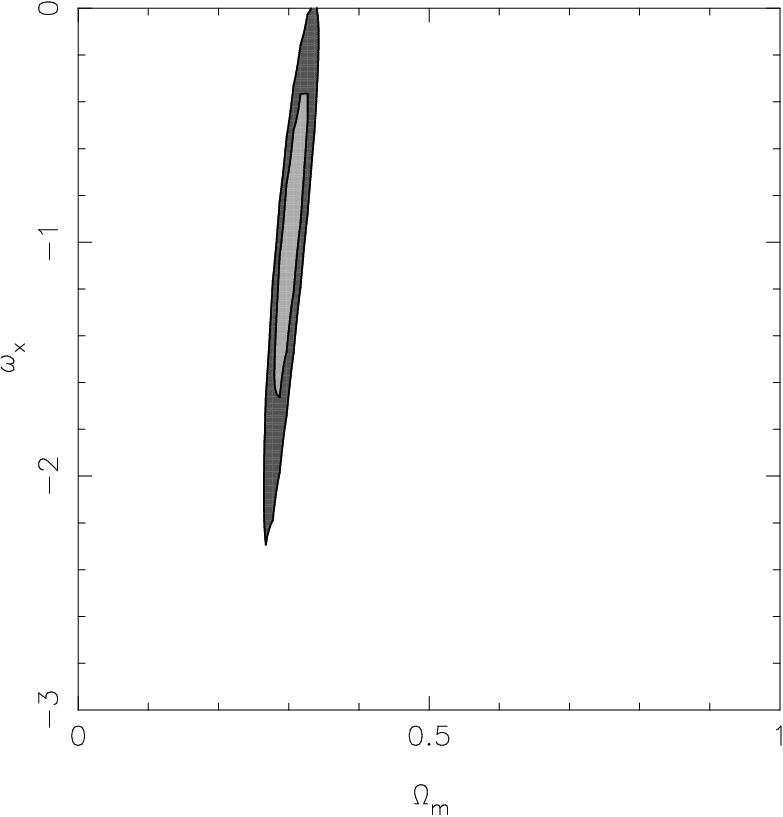}
   \caption{Confidence region plot of the best fit to the $f_{\rm gas}$ of
            9 clusters published by Allen et al. (2002,2003) --
            see the text for a detailed description of the method.
        The 68\% and 95\% confidence levels in the $\omega_x$--$\Omega_m$ plane
            are shown in lower shaded and lower $+$ darker shaded areas
            respectively.
           }
   \label{Fig_cont2}
   \end{figure*}

Using the {\it Chandra} observational data, Allen et al. (2002; 2003) have
  got the $f_{\rm gas}$ profiles for the 10 relaxed clusters.
Except for Abell 963, the $f_{\rm gas}$ profiles of the other 9 clusters
  appear to have converged or be close to converging with a canonical radius
  $r_{2500}$, which is defined as the radius within which the mean mass 
  density is 2500 times the critical density of the universe at the redshift
  of the cluster (Allen et al. 2002, 2003).
The gas mass fraction values of these nine clusters at $r_{2500}$ (or at the
  outermost radii studied for PKS0745-191 and Abell 478) 
  were shown in Figure~5 of Allen et al. (2003).	
We will use this database to constrain the equation of state of the dark energy
  component, $\omega_x$.
Our analysis of the present data is very similar to the one performed by
  Lima et al. (2003).
However, in additional to including new data from Allen et al. (2003),
  we also take into account the bias between the baryon fractions in galaxy
  clusters and in the universe as a whole.
Following Allen et al. (2002), we have the model function as
\begin{equation}
f_{\rm gas}^{\rm mod}(z_i;\omega_x, \Omega_m) =
      \frac{ b \Omega_b}{\left(1+0.19{h}^{1/2}\right) \Omega_m}
  \left[{h\over 0.5}
	\frac{D^A_{\rm{SCDM}}(z_i)}{D^A(z_i;\omega_x, \Omega_m)}
		\right]^{3/2}
\end{equation}
where the bias factor $b \simeq 0.93$ (Bialek et al. 2001; Allen et al. 2003)
  is a parameter motivated by gas dynamical simulations, which suggest that
  the baryon fraction in clusters is slightly depressed with respect to the
  Universe as a whole 
	(Cen and Ostriker 1994; 
	Eke, Navarro and Frenk 1998;
	Frenk et al. 1999; 
	Bialek et al. 2001).
The term $(h/0.5)^{3/2}$ represents the change in the Hubble parameter from
  the defaut value of $H_0 = 50 {\rm{km \, s^{-1} \, Mpc^{-1}}}$ and
  the ratio $D^A_{\rm{SCDM}}(z_i)/D^A(z_i;\omega_x, \Omega_m)$
  accounts for the deviations of the model considering from the default
  standard cold dark matter (SCDM) cosmology.

Again, we determine $\omega_x$ and $\Omega_m$ through
  a $\chi^{2}$ minimization method with the same parameter ranges and
  steps as last section.
We constrain $\Omega_m h^{2} = 0.0205 \pm 0.0018$, the bound from the
  primodial nucleosynthesis (O'Meara et al. 2001),
  and $h = 0.72 \pm 0.08$, the final result from the Hubble Key Project by
  Freedman et al. (2001).
The $\chi^2$ difference between the model function and SCDM data is then
  (Allen et al. 2003)
\begin{equation}
\label{eq:chi2}
\chi^{2}(\omega_x, \Omega_m) =    \sum_{i = 1}^{9}
 \frac{\left[f_{\rm gas}^{\rm mod}(z_i;\omega_x,\Omega_m)-f_{{\rm gas,o}i}\right]^2}
	{\sigma_{f_{{\rm gas},i}}^2}	+
  \left[\frac{\Omega_bh^{2} - 0.0205}{0.0018}\right]^{2}   +
  \left[\frac{h - 0.72}{0.08}\right]^{2},
\end{equation}
where $f_{\rm gas}^{\rm mod}(z_i;\omega_x,\Omega_m)$ refers to equation~(3),
  $f_{{\rm gas,o}i}$ is the measured $f_{\rm gas}$ with the defaut SCDM
  cosmology, and $\sigma_{f_{{\rm gas},i}}$ is the symmetric root-mean-square
  errors ($i$ refers to the $i$th data point, with totally 9 data).
The summation is over all of the observational data points.

Figure~3 displays the 68.3\% and 95.4\% confidence level contours in the
  ($\omega_x$, $\Omega_m$) plane of our analysis 
  using the lower shaded and the lower plus darker shaded areas respectively.
The best fit happans at $\omega_x = -0.86$ and $\Omega_m = 0.30$ .
As shown in the figure, although the X-ray gas mass fraction data 
  constrains the density parameter $\Omega_m$ very stringently,
  it still poorly limits the dark energy equation of state $\omega_x$. 
The situation can be dramatically improved when the two databases are combined
  to analysis, in particularly, a nontrivial lower bound on $\omega_x$ will
  be obtained (see below).


\section{Combined analysis, conclusion and discussion}

Now we present our combined analysis of the constraints from 
  the angular size-redshift data and the X-ray gas mass fraction of 
  galaxy clusters and summarize our results. 
In Figure~4, we display the 68.3\% and 95.4\% confidence level
  contours in the ($\omega_x$, $\Omega_m$) plane using the lower shaded
  and the lower plus darker shaded areas respectively.
The best fit happans at $\omega_x = -1.16$ and $\Omega_m = 0.29$.
As it shown, 
  fairly stringent bounds on both $\omega_x$ and $\Omega_m$ are obtained,
  with $-2.22< \omega_x< -0.62$ and $0.28< \Omega_m < 0.32$ at the 95.4\% 
  confidence level. The bound on $\omega_x$ goes to $-1 \le \omega_x< -0.60$
  when the constraint $\omega_x \ge -1$ is imposed.

   \begin{figure*}
   \centering
   \includegraphics[width=8.0cm]{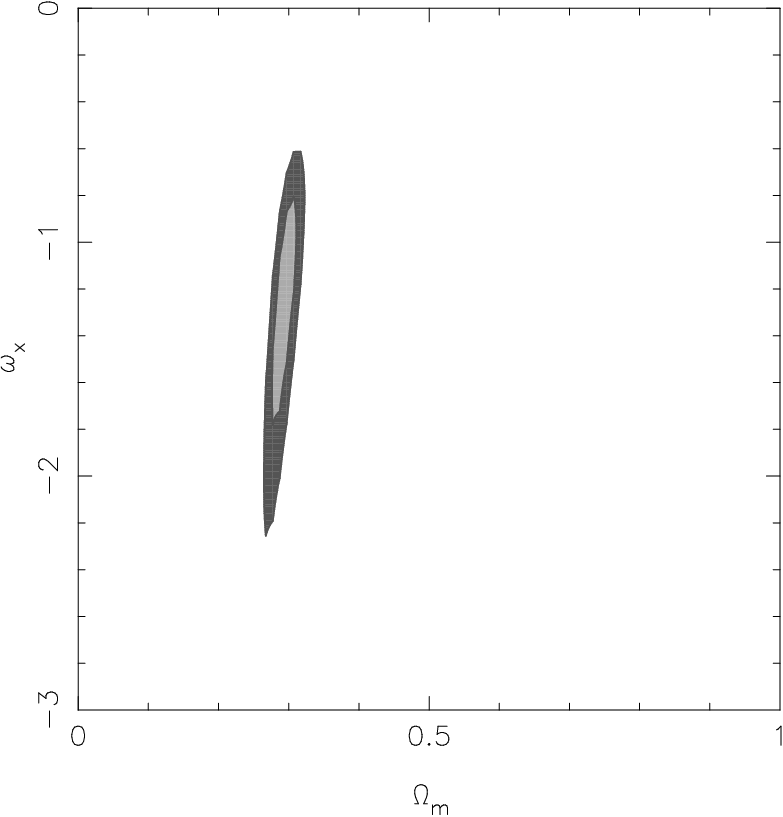}
   \caption{Confidence region plot of the best fit from a combined analysis
	for the angular size-redshift data (Gurvits et al. 1999) and
	the X-ray gas mass fractions of 9 clusters (Allen et al. 2002, 2003).
	The 68\% and 95.4\% confidence levels in the
	$\omega_x$--$\Omega_m$ plane are shown in lower shaded
	and lower + darker shaded areas respectively.
	The best fit happans at $\omega_x = -1.16$ and $\Omega_m = 0.29$.
           }
   \label{Fig_cont}
    \end{figure*}

Although precise determinations of $\omega_x$ and its possible evolution with
  cosmic time are crucial for deciphering the mystery of DE, 
  currently $\omega_x$ has not been determined quite well 
  even with an assumption of $\omega_x$ being constant
	(Hannestad and M\"ortsell 2002;
	Spergel et al. 2003;
	Takahashi et al. 2003). 
It is worthy of determining $\omega_x$ using a joint analysis.  
In this paper we have shown that stringent constraints on $\omega_x$ can
  be obtained from the combination analysis of the angular size-redshift
  data and the X-ray mass fraction data of clusters, which is a complementary
  to other joint analyses.
{
At this point we compare our results with other recent determinations of
  $\omega_x$ from independent methods.
For the usual quintessence model (i.e., the constraint $\omega_x \ge -1$ 
  is imposed), Garnavich et al. (1998) found $\omega_x < -0.55$ using the SNeIa
  data from the High-$z$ Supernova Search Team, while Lima and Alcaniz (2002)
  obtained $\omega_x < -0.50$ using the anugular size-redshift data from
  Gurvits, Kellerman and Frey (1999) (95\% confidence level).
Our result of $\omega_x < -0.60$ is a little bit more stringent than theirs.
However Bean and Melchiorri (2002) found an even better constraint,
  $\omega_x < -0.85$, by analyzing SNeIa data and measurements of LSS and 
  the positions of the acoustic peaks in the CMB spectrum.
For the more general dark energy model including either normal XCDM, as well
  as the extended or phantom energy (i.e., the constraint $\omega_x \ge -1$
  is relaxed), Hannestad and M\"ortsell (2002) combined CMB, LSS and SNeIa data
  for analyzing and obtained $-2.68 < \omega_x < -0.78$ at 95.4\% confidence 
  level, whose lower and upper bounds are a little bit lower than ours
  ($-2.22 < \omega_x < -0.62$ at 95.4\% confidence level).
Recently, Schuecker et al. (2003) combined REFLEX X-ray clusters and SNeIa data 
  to obtain $-1.30 < \omega_x < -0.65$ with $1\sigma$ statistical significance.
From Figure~4, it is found our $1\sigma$ result is $-1.72 < \omega_x < -0.83$,
  which is comparable with the results of Schuecker et al. (2003). 
Using the X-ray gas mass fraction of 6 galaxy clusters, Lima et al. (2003)
  found $-2.08 < \omega_x < -0.60$ ($1\sigma$ level), which is less stringent 
  than the result presented in this work.
This is because we used more X-ray gas mass fraction data of galaxy clusters
 and combined the angular size-redshift data of compact radio sources 
  for analyzing.
}
The analysis presented here reinforces the interest in precise 
  measurements of angular size of distant compact radio sources and statistical
  studies of the intrinsic length distribution of the sources.
It is also hopefully that our constraints will be dramatically improved after
  more acurate X-ray data from {\it Chandra} and {\it XMM-Newton} become
  available near future.

\begin{acknowledgements}
We would like to thank 
  L. I. Gurvits for sending us their compilation of the angular size-redshift
    data and helpful explanation of the data, 
  S. Allen for providing us the X-ray mass fraction data and various help
    about data analysis,
	J. S. Alcaniz
	and D. Tatsumi for their helpful discussion.
{
Our thanks go to the anonymouse referee for valuable comments and useful
  suggestions, which improved this work very much.
}
This work was supported by
  a Grant-in-Aid for Scientific Research on Priority Areas (No.14047219) from
  the Ministry of Education, Culture, Sports, Science and Technology.

\end{acknowledgements}


\begin{thebibliography}{}


\bibitem[Alcaniz 2002]{alc02}
     Alcaniz, J. S. 2002, \prd, 65, 123514   

\bibitem[2002]{}Allen S.W., Schmidt R.W., Fabian A.C., 2002, MNRAS, 334, L11

\bibitem[2003]{all03}Allen S.W., Schmidt R.W., Fabian A.C., Ebeling, H. 2003, 
	MNRAS, 342, 287

\bibitem[Balbi et al. 2000]{bal00}
        Balbi, A. et al. 2000, \apjl, 545, L1

\bibitem[Barris et al. 2003]{bar03}
	Barris, B. J. et al. 2003, \apj, accepted (astro-ph/0310843)

\bibitem{}
	Bean, R. and Melchiorri, A. 2002, \prd, 65, 041302(R)

\bibitem[Bennett et al. 2003]{ben03}
        Bennett, C. L. et al. 2003 \apj, accepted (astro-ph/0302207)

\bibitem{} 
	Bialek J.J., Evrard A.E., Mohr J.J., 2001, ApJ, 555, 597

\bibitem{} 
	Biesiada, M. 2001, \mnras, 325, 1075

\bibitem{}
	Caldwell, R. R. 2002, Phys.Lett.B, 545, 23

\bibitem[Caldwell et al. 1988]{cal98}
        Caldwell, R., Dave, R., and Steinhardt, P. J. 1998, \prl, 80, 1582

\bibitem[Carroll et al. 1992]{car92}
        Carroll, S., Press, W. H. and Turner, E. L. 1992, \araa, 30, 499

\bibitem{} Cen R., Ostriker J.P., 1994, ApJ, 429, 4

\bibitem{chae}
        Chae, K.-H. et al. 2002, Phys. Rev. Lett., 89, 151301

\bibitem[Chen and Ratra (2003)]{che03}
	Chen, G., and Ratra, B. 2003, \apj, 582, 586

\bibitem[Chiba et al. 1997]{chi97}
        Chiba, T., Sugiyama, N. and Nakamura, T. 1997, \mnras, 289, L5

\bibitem{} 
	Choubey, S. and King, S. F. 2003, \prd, 67, 073005

\bibitem[de Bernardis et al. 2000]{ber00}
        de Bernardis, P. et al. 2000, \nat, 404, 955

\bibitem[Dev et al. 2003]{dev03}
	Dev, A., Jain, D. and Mahajan, S. 2003, astro-ph/0307441

\bibitem[Durrer et al. 2003]{dur03}
        Durrer, R., Novosyadlyj, B. and Apunevych, S. 2003, \apj, 583, 33

\bibitem{}
	Efstathiou G. 1999, MNRAS, 310, 842

\bibitem{} Eke V.R., Navarro J.F., Frenk C.S., 1998, ApJ, 503, 569

\bibitem{}
	Frampton, P. H. 2003, Phys.Lett.B, 555, 139	

\bibitem{} 
	Freedman W. et al., 2001, ApJ, 553, 47

\bibitem{} 
	Frenk C.S. et al., 1999, ApJ, 525, 554

\bibitem{}
	Garnavich P. M. et al. 1998, ApJ, 509, 74

\bibitem{} 
	Gurvits, L. I., Kellerman, K. I. and Frey, S. 1999, \aap, 342, 378

\bibitem{}
	Haiman, Z. Mohr, J. J., and Holder, G. P. 2001, \apj, 553, 545

\bibitem{}
	Hannestad, S. and M\"ortsell, E. \prd, 66, 063508

\bibitem{}
	Huterer, D. and Ma, C. -P. 2003, \apjl, submitted (astro-ph/0307301)

\bibitem{}
	Huterer, D. and Turner, M. S. 1999, \prd, 60, 081301

\bibitem[Jain et al. 2003]{jai03}
        Jain, D., Dev, A. and Alcaniz, J. S. 2003, Class. Quan. Grav. 20, 4163

\bibitem{}
	Jimenez, J., Verde, L., Treu, T., Stern, D. 2003, \apj, submitted
	(astro-ph/0302560)

\bibitem[Knop et al. 2003]{kno03}
	Knop, R. A. et al. 2003, \apj, accepted (astro-ph/0309368)

\bibitem{}
	Kujat, J., Linn, A. M., Scherrer, R. J., Weinberg, D. H.
	2002, \apj, 572, 1

\bibitem{}
	Lima J. A. S. \& Alcaniz J. S. 2000a, MNRAS, 317, 893

\bibitem{}
        Lima J. A. S. \& Alcaniz J. S. 2000b, \aap, 357, 393

\bibitem{}
        Lima J. A. S. \& Alcaniz J. S. 2002, \apj, 566, 15

\bibitem[Lima et al. 2003]{lim03}
       Lima, J. A. S., Cunha, J. V. and Alcaniz, J. S. 2003, \prd, 68, 023510

\bibitem{}
	Maor, I. Brustein, R., McMahon, J. and Steinhardt, P. J. 2002, 
	\prd, 65, 123003

\bibitem{}
	Maor, I. Brustein, R. and Steinhardt, P. J. 2001, \prl, 86, 6

\bibitem{}
	Melchiorri, A., Mersini, L. \"Odman, C. J. Trodden, M. 2003, 
	\prd, 68, 043509

\bibitem{} 
	O'Meara J.M., Tytler D., Kirkman D., Suzuki N., Prochaska J.X., 
	Lubin D., Wolfe A.M., 2001, ApJ, 552, 718

\bibitem{} 
	Parker, L. and Raval, A. 1999, \prd, 60, 063512

\bibitem{} Perlmutter, S. et al. 1998, \nat, 391, 51

\bibitem[Perlmutter et al. 1999]{per99}
        Perlmutter, S. et al. 1999, \apj, 517, 565

\bibitem{}
	Perlmutter S., Turner M. S. \& White M. 1999, \prl, 83, 670

\bibitem[Ratra and  Peebles 1988]{rat88}
        Ratra, B. and P.J.E. Peebles, P. J. E. 1988, \prd, 37, 3406

\bibitem[Riess et al. 1998]{rie98}
        Riess, A. G. et al. 1998, \aj, 116, 1009

\bibitem[Riess et al. 2001]{rie01}
        Riess, A. G. et al. 2001, \apj, 560, 49

\bibitem[]{} Sasaki, S. 1996, PASJ, 48, L119

\bibitem{}
	Schuecker, P., Caldwell, R. R., Böhringer, H., Collins, C. A., 
	Guzzo, L., Weinberg, N. N. 2003, \aap, 402, 53

\bibitem{}
	Schulz, A. E. and White, M. J. 2002, \prd, 64, 043514

\bibitem[Sereno 2002]{ser02}
        Sereno, M. 2002, \aap, 393, 757

\bibitem[Spergel et al. 2003]{spe03}
        Spergel, D. N. et al. 2003 \apj, accepted (astro-ph/0302209)

\bibitem{}
	Steinhardt, P. J. Wang, L. and Zlatev, I. 1999, \prd, 59, 123504

\bibitem{}
	Takahashi, K., Oguri, M., Kotake, K., Ohno, H. 2003, astro-ph/0305260

\bibitem{}
	Tonry, J. L. et al. 2003, \apj, accepted (astro-ph/0305008)

\bibitem[Turner 2002]{tur02}
        Turner, M. S. 2002, \apjl, 576, L101

\bibitem[Turner and White 1997]{tur97}
        Turner, M. S. and White, M. 1997, \prd, 56, R4439  

\bibitem[Vishwakarma 2001]{vis01}
        Vishwakarma, R. G. 2001, Class.Quan.Grav. 18, 1159  

\bibitem{}
	Waga I. \&  Miceli A. P. M. R. 1999, \prd, 59, 103507

\bibitem{}
	Wang, L. Caldwell, R. R. Ostriker, J. P., Steinhardt, P. J. 2000,
	\apj, 530, 17

\bibitem{}
	Wasserman, I. 2002, \prd, 66, 123511

\bibitem[Weinberg 1989]{wei89}
        Weinberg, S. 1989, Rev. Mod. Phys. 61, 1

\bibitem{}
	Weller, J. and Albrecht, A. 2001, \prl, 86, 1939

\bibitem{}
	Weller, J. and Albrecht, A. 2002, \prd, 65, 103512

\bibitem[Wetterich 1988]{wet88}
        Wetterich, C. 1988, Nucl.Phys. B302, 645

\bibitem{} 
	White S.D.M., Navarro J.F., Evrard A.E., Frenk C.S., 1993, 
	Nature, 366, 429

\bibitem[Zhu 1998]{zhu98}
     Zhu, Z. -H. 1998 \aap, 338, 777


\bibitem[Zhu and Fujimoto 2002]{zhu02}
        Zhu, Z. -H. and Fujimoto, M. -K. 2002, \apj, 581, 1

\bibitem[Zhu and Fujimoto 2003]{zhu03}
        Zhu, Z. -H. and Fujimoto, M. -K. 2003, \apj, 585, 52


\bibitem[Zhu, Fujimoto and Tatsumi 2001]{zhu01}
        Zhu, Z. -H., Fujimoto, M. -K. and Tatsumi, D. 2001, \aap, 372, 377

\bibitem[Zlatev et al. 1999]{zla99}
        Zlatev, I., Wang, L. and Steinhardt, P. J. 1999, \prl, 82, 896

\end{thebibliography}
\end{document}